\documentclass[referee]{raa}            
\usepackage{graphicx,times}             
\input{epsf.sty}                        
\input{psfig.sty}                       

\begin{document}

   \title{Interaction between Granulation and Small-Scale Magnetic Flux Observed by Hinode}

   \volnopage{Vol.0 (200x) No.0, 000--000}      
   \setcounter{page}{1}          

   \author{Jun Zhang
      \inst{}
   \and Shuhong Yang
      \inst{}
   \and Chunlan Jin
      \inst{}
   }

   \institute{Key Laboratory of Solar Activity, National Astronomical Observatories, Chinese Academy of
Sciences, Beijing 100012, China; {\it zjun@ourstar.bao.ac.cn}
   }

   \date{Received~~2009 month day; accepted~~2009~~month day}

\abstract{With the polarimetric observations obtained by the
Spectro-Polarimeter on board Hinode, we study the relationship
between granular development and magnetic field evolution in the
quiet Sun. 6 typical cases are displayed to exhibit interaction
between granules and magnetic elements, and we have obtained the
following results. (1) A granule develops centrosymmetrically when
no magnetic flux emerges within the granular cell. (2) A granule
develops and splits noncentrosymmetrically while flux emerges at
an outer part of the granular cell. (3) Magnetic flux emergence as
a cluster of mixed polarities is detected at the position of a
granule as soon as the granule breaks up. (4) A dipole emerges
accompanying with the development of a granule, and the two
elements of the dipole root in the adjacent intergranular lanes
and face each other across the granule. Advected by the horizontal
granular motion, the positive element of the dipole then cancels
with pre-existing negative flux. (5) Flux cancellation also takes
place between a positive element, which is advected by granular
flow, and its surrounding negative flux. (6) While magnetic flux
cancellation takes place at a granular cell, the granule shrinks
and then disappears. (7) Horizontal magnetic fields enhance at the
places where dipoles emerge and where opposite polarities cancel
with each other, but only the horizontal fields between the
dipolar elements point orderly from the positive element to the
negative one. Our results reveal that granules and small-scale
magnetic flux influence each other. Granular flow advects magnetic
flux, and magnetic flux evolution suppresses granular development.
There exist extremely large Doppler blue-shifts at the site of one
cancelling magnetic element. This phenomenon may be caused by the
upward flow produced by magnetic reconnection below the
photosphere. \keywords{Sun: granulation --- Sun: magnetic fields
--- Sun: photosphere --- techniques: polarimetric} }

   \authorrunning{J. Zhang, S.-H. Yang \& C.-L. Jin }            
   \titlerunning{Interaction between Granulation and Magnetic Flux}  

   \maketitle
%
%
\section{Introduction}           
\label{sect:intro}

Both observations and simulations reveal that granules and
small-scale magnetic elements spread all over the quiet Sun.
Granules are bright isolated elements surrounded by intergranular
lanes. The mean size of granules excluding the surrounding dark
lanes amounts to 1$''$.1 (Namba \& Diemel 1969), or 1$''$.35 (Bray
et al. 1984), while the mean cell size of the granular elements
including one-half of the surrounding dark lanes is 1$''$.94 (Bray
\& Loughhead 1977), or 1$''$.76 (Roudier \& Muller 1986).
Frequently the granules expand and split into smaller components
that drift apart, and the fragments may turn grow and fragment, or
merge with others, or shrink and decompose.

Magnetic fields in the quiet Sun can be classified into three
categories based on their locations and morphologies: network
(Leighton et al. 1962), intranetwork (IN; Livingston \& Harvey
1975) and ephemeral regions (Harvey \& Martin 1973). The network
elements are confined to the supergranular boundaries and the IN
ones locate within the supergranular cells. The spatial
distribution and time evolution of IN magnetic features are
closely associated with the solar granulation (Lin \& Rimmele
1999). Small flux and size with rapid time changes make the IN
field difficult to observe and characterize (Keller et al. 1994).
However, much progress has been made in IN morphology dynamics and
some quantitative aspects, such as flux distribution (Wang et al.
1995), lifetime (Zhang et al. 1998a), mean horizontal velocity
field (Wang et al. 1996; Zhang et al. 1998b), motion pattern and
evolution (Zhang et al. 1998b, c, 2006).

The magnetic flux emergence seems to be significantly influenced
by the granular motion. The research about the horizontal IN
fields suggests that small magnetic loops are being advected
toward the surface by the convective upward motion of the plasma
inside the granules (Lites, et al. 1996; Orozco Su{\'a}rez et al.
2008). The horizontal motion inside the granules carries the
vertical magnetic flux toward the intergranular lanes (Harvey et
al. 2007; Centeno et al. 2007). Then most of these IN magnetic
elements are destroyed by three mechanisms: merging with IN or
network elements of the same polarity, cancellation of opposite
polarity elements, or separation and disappearance at the position
where they appear (Zhang et al. 1998a). Furthermore, the magnetic
emergence also has important influence on the shape of the
underlying granulation pattern leading to the so-called ``abnormal
granulation" (Cheung et al. 2007).

Using the continuum intensities, vector magnetic fields and
Doppler velocities derived from the Stokes profiles obtained by
the Spectro-Polarimeter (SP; Lites et al. 2001) on board Hinode
(Kosugi et al. 2007), we mainly study the relationship between the
development of granular structures and the emergence and
cancellation of small-scale (with a typical size of $\sim$1$''$ in
this paper) magnetic elements from their birth to death. In Sect.
2, we describe the observations and the strategy of Stokes profile
inversion. Then we present the relationship between granular
development and magnetic flux emergence (in Sect. 3) and
cancellation (in Sect. 4). The conclusions and discussion are all
given in Sect. 5.

\section{Observations and inversion strategy}
\label{sect:Obs}

\begin{figure}
\centering
\includegraphics[bb=185 84 386
741,clip,angle=0,scale=.82]{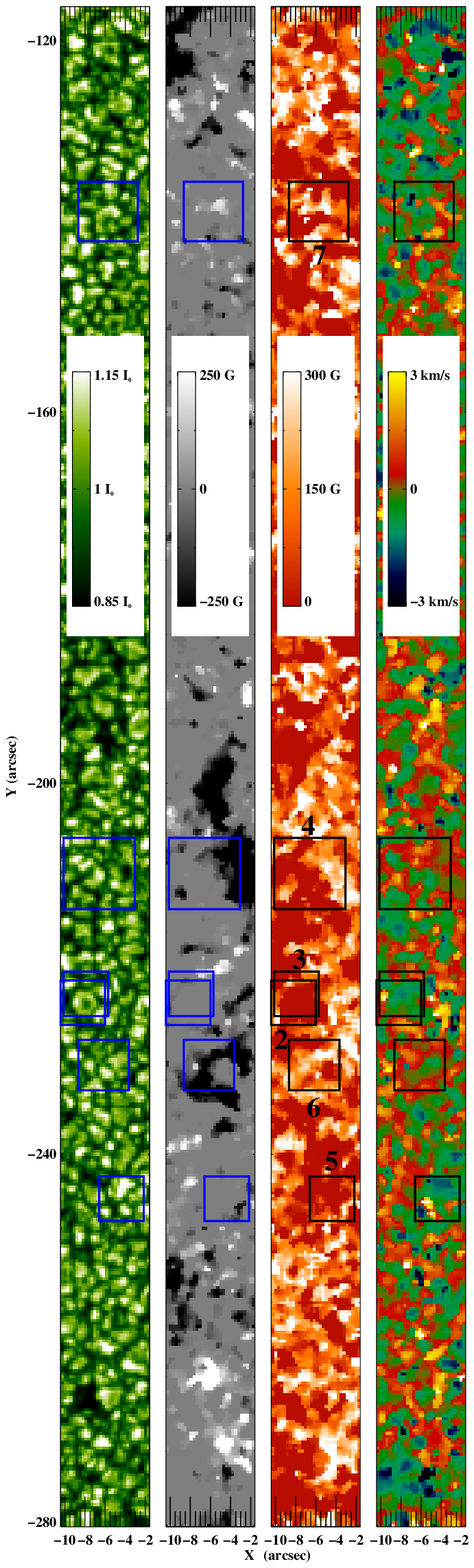} \caption{General
appearance of the whole field-of-view (FOV) retrieved from the SP
data obtained from 14:29:54 UT to 14:31:51 UT on June 1, 2007.
{\it From left to right}: continuum intensity map, corresponding
longitudinal magnetogram, transverse magnetogram, and Dopplergram.
Windows 2$-$7 outline the FOVs of Figs. 2$-$7, respectively.
I$_{0}$ represents the average continuum intensity in the whole
FOV. \label{fig1}}
\end{figure}

The SP instrument of the Solar Optical Telescope (SOT; Ichimoto et
al. 2008; Shimizu et al. 2008; Suematsu et al. 2008; Tsuneta et
al. 2008) aboard Hinode provides the full Stokes profiles of two
Fe lines at 630.15 nm ($g_{eff}=1.67$) and 630.25 nm
($g_{eff}=2.5$) in four modes (fast map, dynamics, normal map and
deep magnetogram). In order to investigate both the structure and
the evolution of granules and magnetic elements, the observational
field-of-view (FOV) should not be too small and the cadence should
be high enough. So we adopt the data taken from 11:33 UT to 17:51
UT on June 1, 2007 in the fast map mode. These data consist of 92
sets of SP maps with a 2-minute cadence, and the observational
target is a quiet Sun region near disk center ($-$6$''$,
$-$199$''$) with a FOV of 8$''$.86$\times$162$''$.3. The scan step
(X direction in Fig. 1) is 0$''$.295, and the pixel sampling along
the slit direction (Y direction in Fig. 1) is 0$''$.32. The
integration time for each slit was 3.2 s with a noise level in the
polarization continuum of 1.4$\times$$10^{-3}I_{c}$.

By using the inversion techniques based on the assumption of
Miline$-$Eddington atmosphere model (Yokoyama 2008, in
preparation), we can derive vector magnetic fields from the full
Stokes profiles. Although the inversion procedures encounter
difficulties in convergence toward and uniqueness of the solutions
when confronting with noisy profiles (Lites et al. 2008), they
will be largely independent of the noise and the field strength
initialization if only the pixels with polarization signals above
a reasonable threshold are inverted (Orozco Su\'{a}rez et al.
2007). Here we only analyze the pixels with total polarization
degrees above 1 time of the noise level in the polarization
continuum in order to exclude some profiles that cannot be
inverted reliably.

Values of 13 free parameters are returned from the inversion,
including the three components of magnetic field (field strength
$B$, inclination angle $\gamma $, azimuth angle $\phi$), the stray
light fraction $\alpha$, the Doppler velocity $V_{los}$, and so
on. Since the pair of Fe I lines in low flux quiet Sun regions are
not capable of distinguishing between the intrinsic magnetic field
and the filling factor (Mart\'{i}nez Gonz\'{a}lez et al. 2006),
the flux density is a more appropriate quantity to describe. Here,
we show the equivalent, spatially resolved vector magnetic fields
by ``apparent flux density" of the longitudinal and transverse
components (Lites et al. 1999), i.e.
$B^{L}_{app}=(1$-$\alpha)Bcos\gamma$ and
$B^{T}_{app}=(1$-$\alpha)^{1/2}Bsin\gamma$, respectively. The
longitudinal component $B^{L}_{app}$ may be considered as the
magnitude of the line-of-sight (LOS) component of a spatially
resolved magnetic field that produces the circular polarization
signal as the observed signal, and the transverse component
$B^{T}_{app}$ is vertical to the LOS that would produce the
observed linear polarization signal. In the vector field
measurements based on the Zeeman Effect, there exists 180 degree
ambiguity in determining the field azimuth. Potential field
approximation is one of the fairly acceptable methods to resolve
the ambiguity (Wang 1999). The Doppler velocities, which are
evaluated from the center of the Stokes I profiles according to Fe
630.25 nm line and averaged over the whole FOV, are well defined
even in weak field regions (Chae et al. 2004).

Figure 1 shows the whole FOV images retrieved from the SP data
obtained from 14:29:54 UT to 14:31:51 UT. They are continuum
intensity map, corresponding longitudinal magnetogram, transverse
magnetogram, and Dopplergram from left to right, respectively. The
zero value areas in the magnetograms represent the pixels not
included in the analysis because of their low polarization
signals. Windows 2$-$7 in Fig. 1 outline the corresponding FOVs of
Figs. 2$-$7, respectively. Windows 4 and 6 locate near network
magnetic fields, while others at IN fields.

\section{Relationship between granular development and magnetic flux emergence}
\label{sect:Result-1}

In order to explore the relationship between granular development
and magnetic flux emergence in the quiet Sun, we examine time
sequence of continuum intensity maps, corresponding vector
magnetograms and Dopplergrams. We find that granules and magnetic
elements influence each other in complicated ways. Here we display
several typical cases representing different interacting forms
between them.

\begin{figure}
\centering
\includegraphics[bb=96 218 526
610,clip,angle=0,scale=1.0]{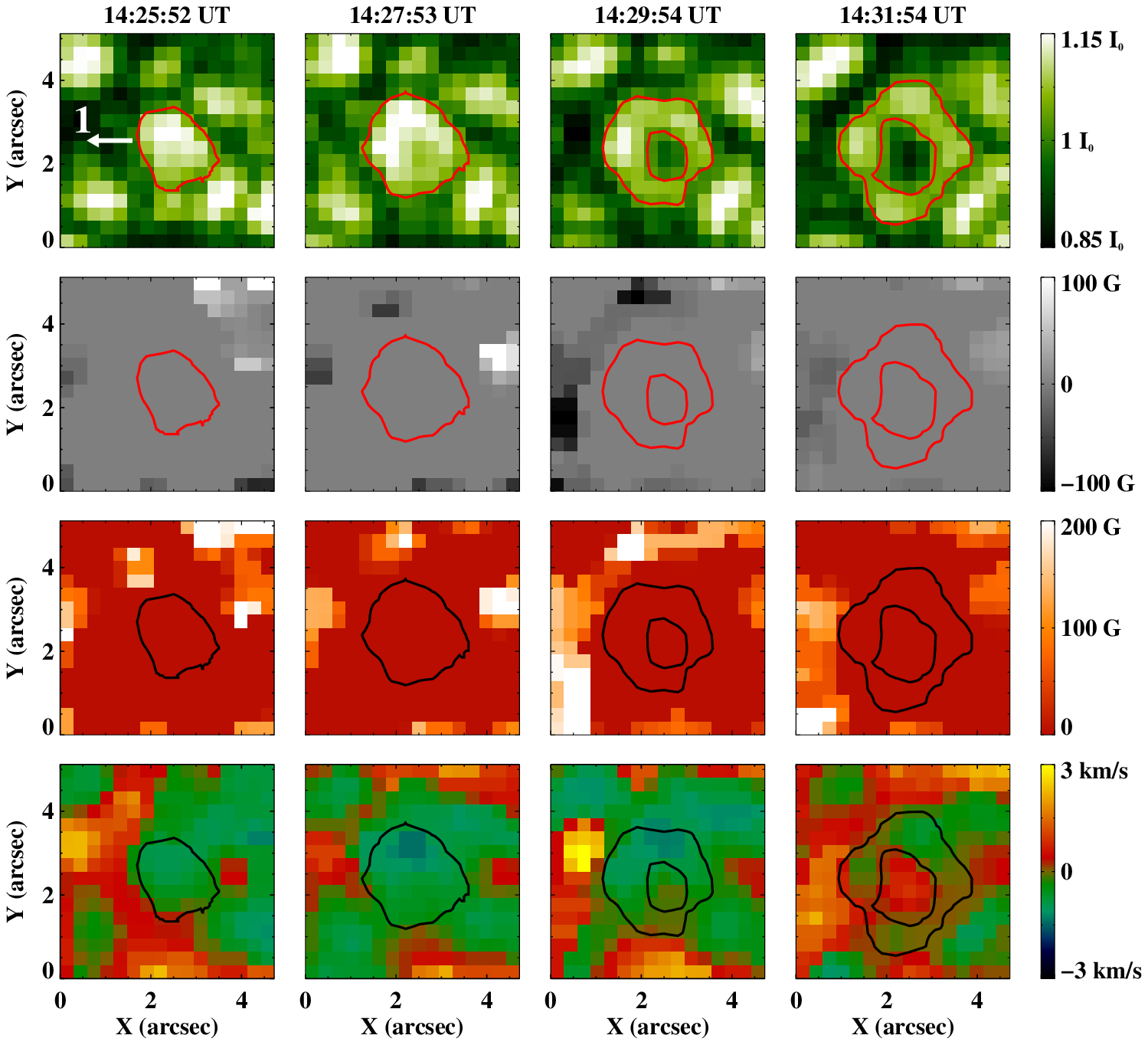} \caption{ Temporal
evolution of a granule with no magnetic flux emergence.
\textit{From top to bottom}: continuum intensity, corresponding
longitudinal field, transverse field and Doppler velocity. The
contours in the top row panels represent the granule with
continuum intensity I$_c$$/$I$_0$$=$1.03, and they are also
overplotted to the corresponding magnetograms and Dopplergrams.
Here, only the granule we focus to study is selected to outline.
Arrow ``1" denotes the direction of the granular flow along which
we calculate the apparent horizontal velocity.
 \label{fig2}}
\end{figure}

It appears that a granular structure develops centrosymmetrically
when no magnetic flux emerges within the granular cell, just as
shown in Fig. 2. The contour curves outline the granule focused to
study with continuum intensity ratio I$_{c}$$/$I$_{0}$$=$1.03,
where I$_0$ represents the average continuum intensity in the
whole FOV. From 14:25 UT on, the granule expanded continuously
with a mean apparent horizontal velocity of 1.5 km s$^{-1}$
(calculated along the direction of arrow ``1"). At 14:29 UT, a
dark core appeared near the granular center and expanded larger.
Comparing the maps of continuum intensity with the Dopplergrams,
we can see that the granular cell always suffered Doppler
blue-shifts during their development process.

\begin{figure}
\centering
\includegraphics[bb=96 218 526
610,clip,angle=0,scale=1.0]{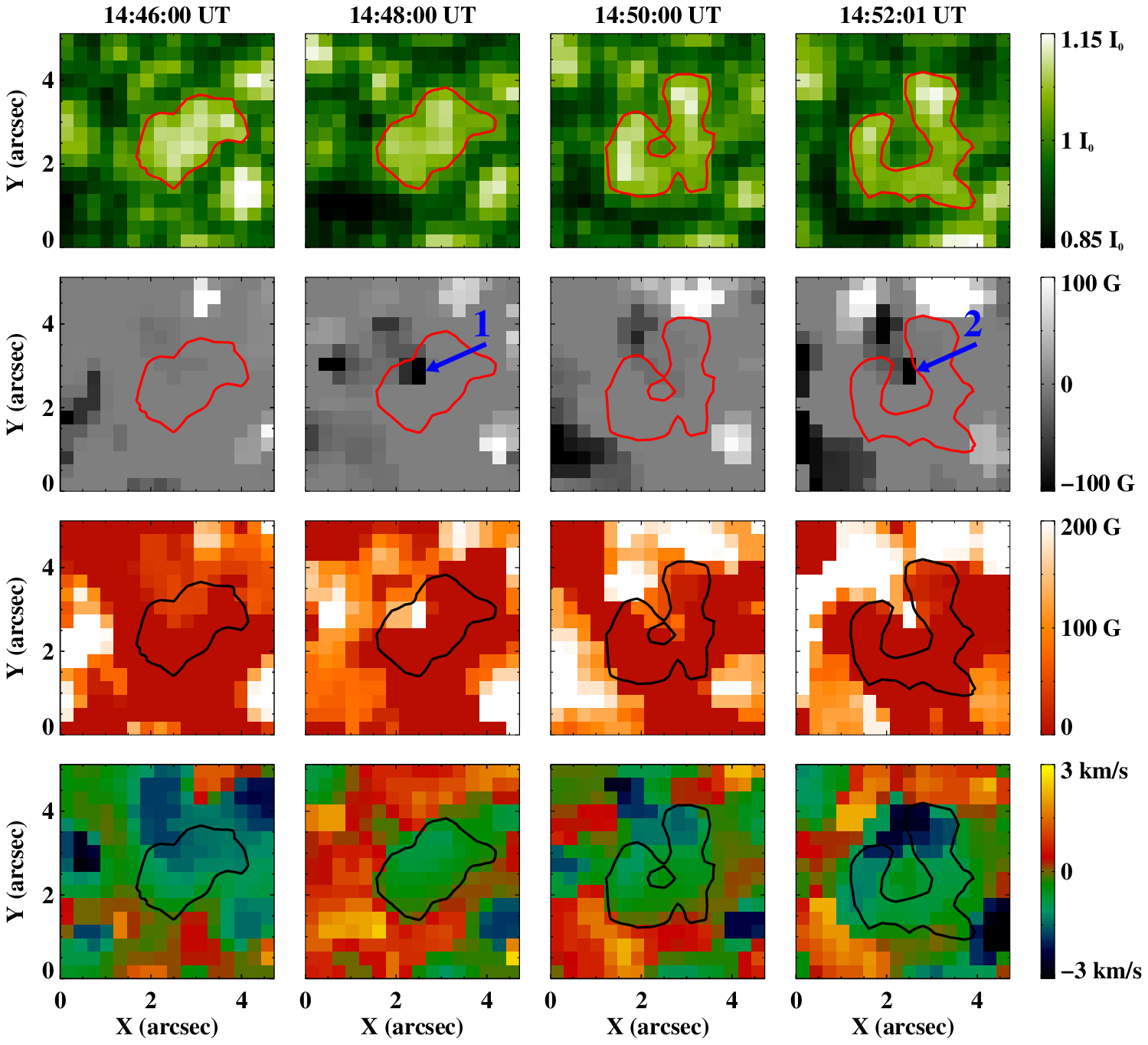} \caption{Similar to
Fig. 2 but for a granule accompanying flux emergence. Arrows ``1"
and ``2" denote two magnetic elements emerging orderly at an outer
part of the granule. \label{fig3}}
\end{figure}

We have also noticed that a granular structure develops and splits
noncentrosymmetrically while magnetic flux emerges within the
granular cell. A typical example is exhibited in Fig. 3. Arrows
``1" and ``2" denote two negative elements emerging orderly at an
outer part of the granule within 6 minutes, and their maximum
longitudinal apparent flux density reached 100 Mx cm$^{-2}$.
Different from the granular shape shown in Fig. 2, the boundary of
this granule became concave as the magnetic flux emerged. At 14:50
UT, the granule split into several small fragments.

\begin{figure}
\centering
\includegraphics[bb=96 260 526
569,clip,angle=0,scale=1.08]{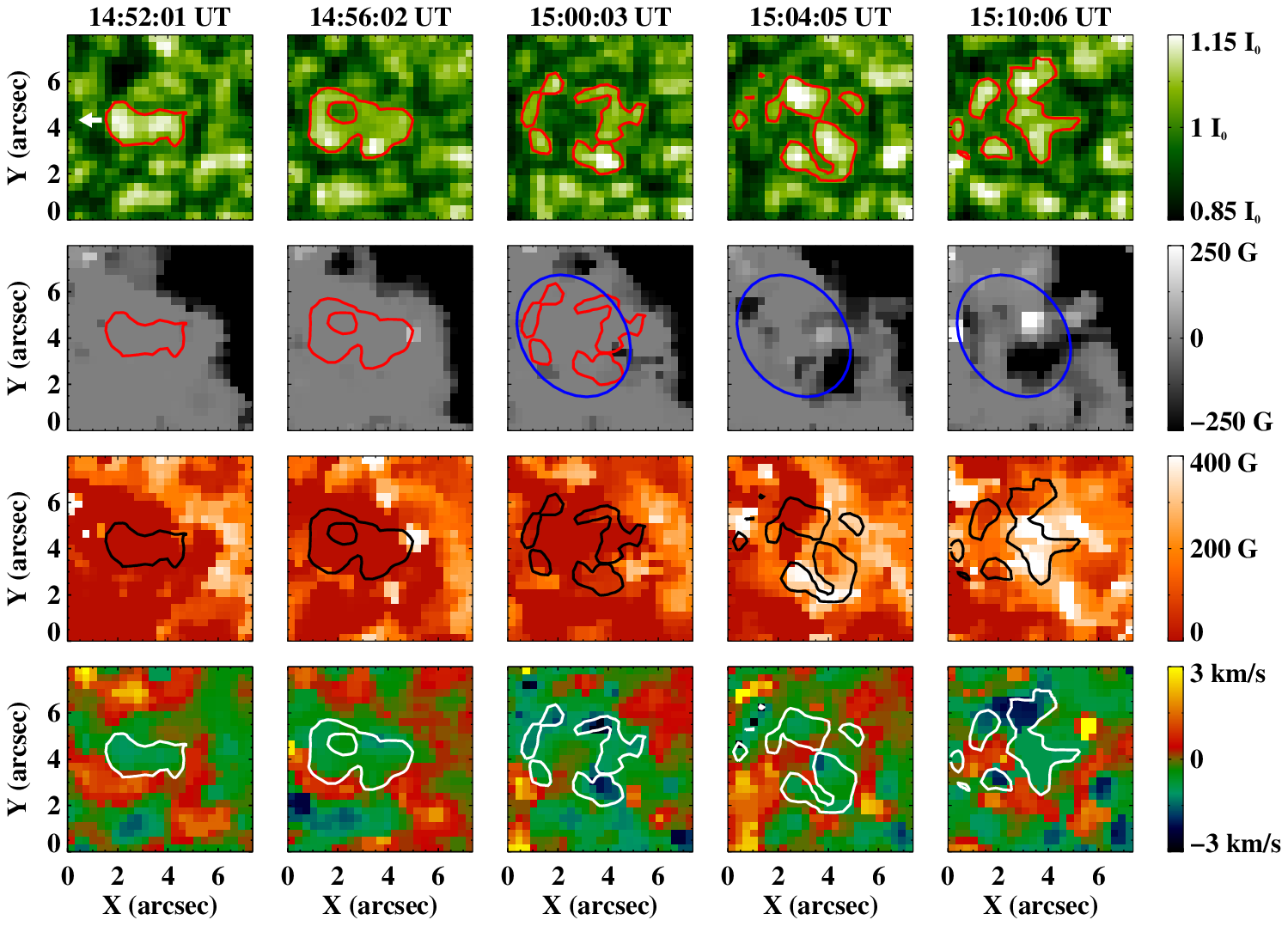} \caption{Similar to
Fig. 2 but for magnetic flux emergence as a cluster of mixed
polarities following the split of a granule. The ellipses enclose
the area where the flux emerges. \label{fig4}}
\end{figure}

Figure 4 shows that lots of magnetic elements emerged as a cluster
of mixed polarities while a granule broke up. The granule first
appeared at 14:46 UT, then developed gradually and became larger
with a mean apparent horizontal expanding velocity of 1.6 km
s$^{-1}$ along the direction indicated by the arrow. The main body
of the granule split at 15:00 UT; meanwhile the magnetic flux was
detected emerging as a cluster of mixed polarities at the position
the initial granule located (defined by the ellipses in the
longitudinal magnetograms), and reached its maximum absolute value
of 4.2$\times$10$^{18}$ Mx at 15:10 UT.

Observations also indicate that the granular motion has
significant influence on the magnetic flux emergence. As shown in
the first two columns of Fig. 5, a dipole emerged accompanying
with the development of a granule. A pair of arrows labelled ``2"
denote the two magnetic elements of the dipole. At 16:12 UT, the
positive element of the dipole appeared and the transverse fields
had generally ordered directions (shown by thin arrows in the
transverse magnetograms at 16:12 UT). At this time, the granular
region suffered larger Doppler blue-shifts with a mean velocity of
$-$1.8 km s$^{-1}$. Two minutes later, the other element of the
dipole appeared and the transverse fields between the dipolar
elements enhanced with their directions pointing orderly from the
positive element to the negative one. The two elements of the
dipole then lay on the two sides of the granule, rooting in the
adjacent intergranular lanes.

\section{Relationship between granular development and magnetic flux cancellation}
\label{sect:Result-2}

Magnetic flux cancellation is a main form of flux disappearance in
the solar photosphere, but its physical mechanism is not clearly
known in detail. As lacking magnetic field observations with high
temp-spatial resolution, the behavior of flux cancellation at a
sub-arcsec spatial scale has not been well researched.

\begin{figure}
\centering
\includegraphics[bb=96 260 526
569,clip,angle=0,scale=1.08]{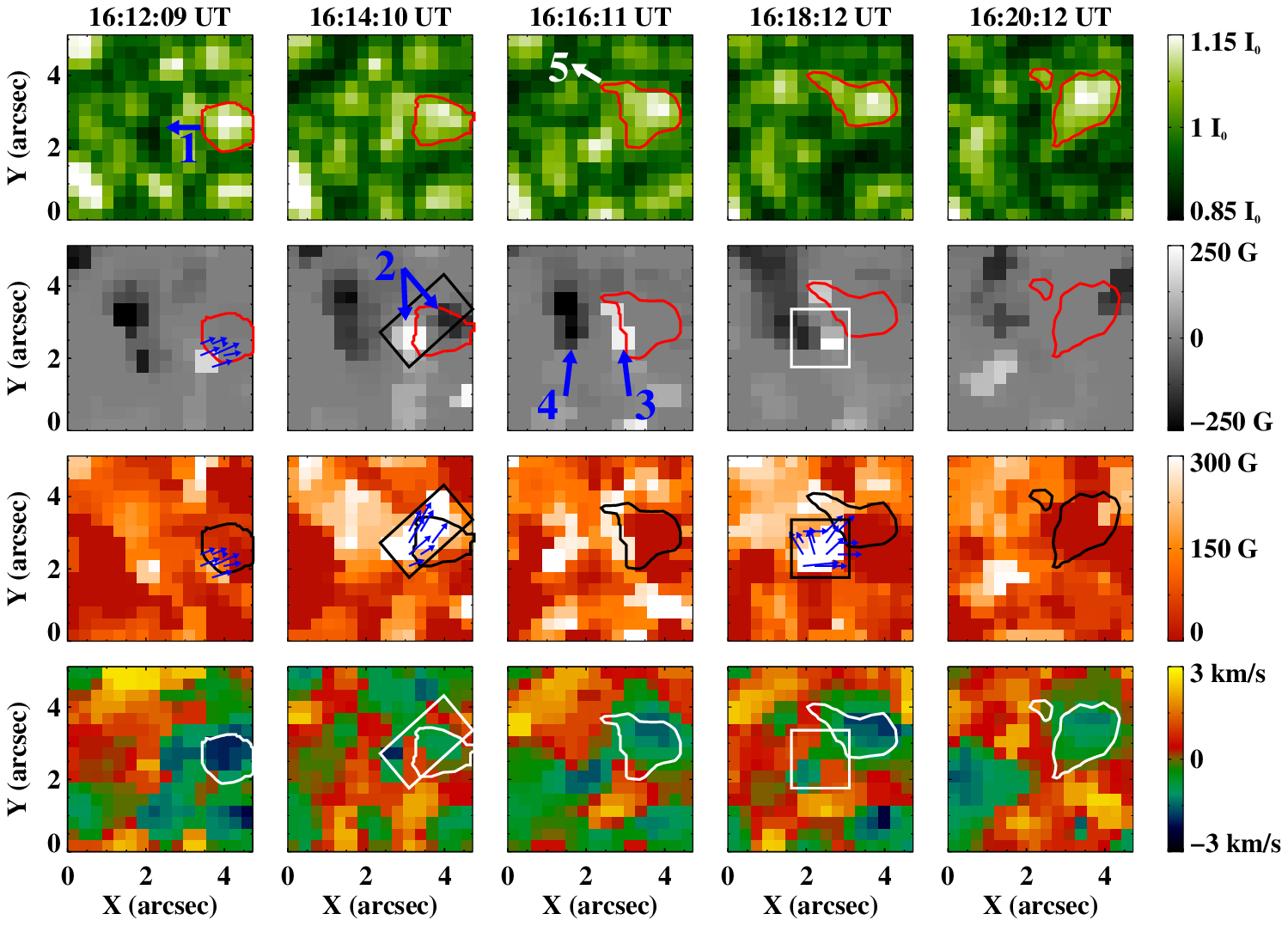} \caption{Similar to
Fig. 2 but for emergence of a dipole (first two columns; marked by
a pair of arrows ``2") and cancellation (last three columns)
between a positive element (indicated by arrow ``3") belonging to
the dipole and a pre-existing negative element (shown with arrow
``4"). The parallelograms and squares embody the regions where the
transverse fields enhance, and the thin arrows indicate the
directions of the local horizontal fields. Arrow ``1" denotes the
direction along which we calculate the apparent horizontal
velocity of the granular flow and arrow ``5" the direction along
which the granule quickly extended. \label{fig5}}
\end{figure}

The last three columns in Fig. 5 exhibit the flux cancellation
which occurred between the positive element of the dipole and
pre-existing negative flux. The positive element (shown by arrow
``3") was advected by the horizontal flow with a mean velocity of
1.7 km s$^{-1}$ along arrow ``1" direction towards the
pre-existing negative element (denoted by arrow ``4"). At 16:18
UT, they encountered each other and cancelled violently; meanwhile
strong transverse fields (see the square area) appeared, but their
directions were unordered, i.e. with no orientation consistency.
By 16:20 UT, the positive element had disappeared completely.

\begin{figure}
\centering
\includegraphics[bb=96 218 526
610,clip,angle=0,scale=1.0]{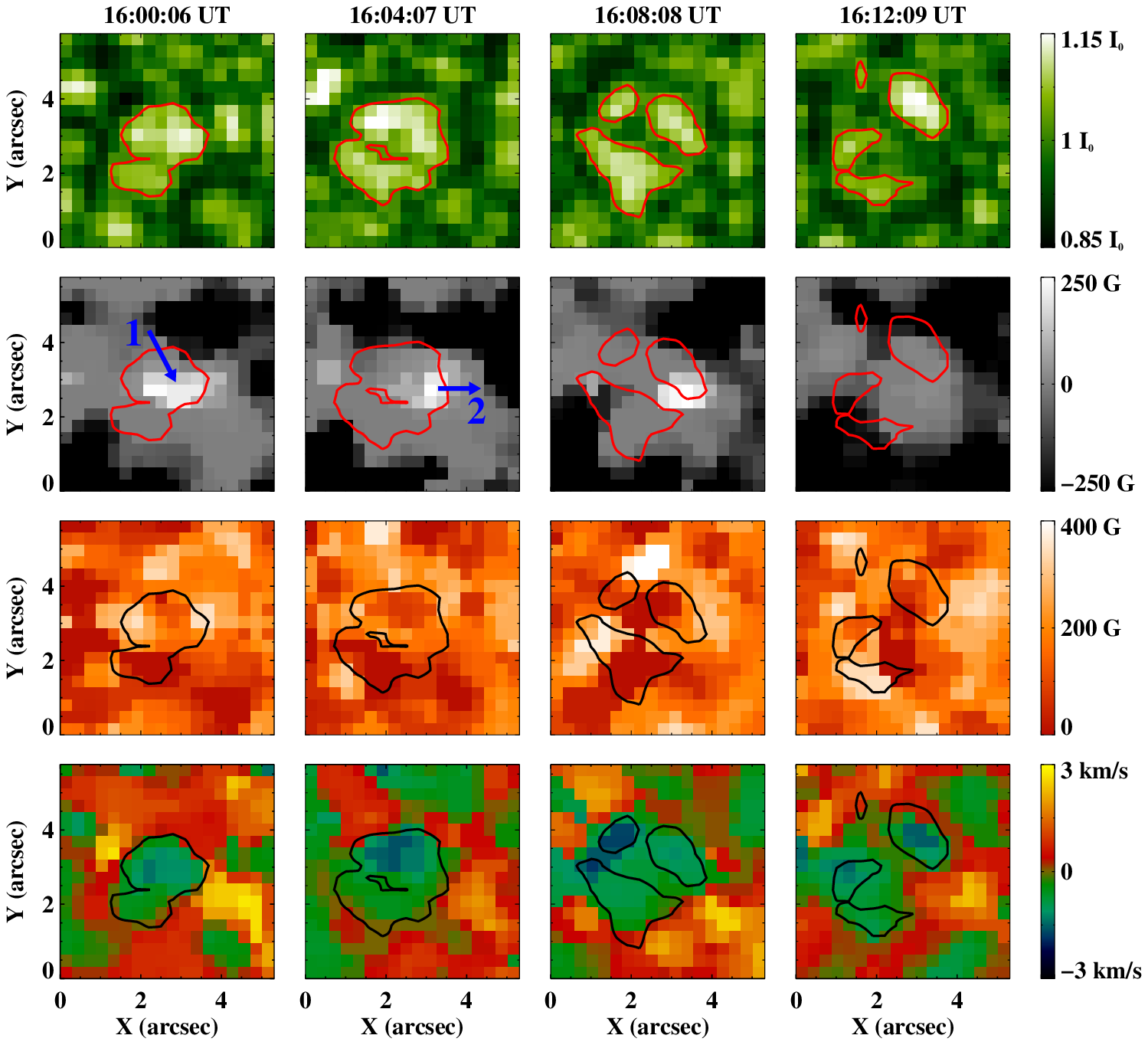} \caption{Similar to
Fig. 2 but for flux cancellation of a positive element advected by
the granular flow with its surrounding negative flux. Arrow ``1"
denotes the positive element and arrow ``2" moving direction of
the element. \label{fig6}}
\end{figure}

Figure 6 displays a case of magnetic flux cancellation between a
positive element (indicated by arrow ``1") and its surrounding
negative flux. During the growing and splitting process of a
granule, the positive element was advected by the granular flow
along arrow ``2" direction with a mean velocity of 1.0 km
s$^{-1}$. At 16:08 UT, it encountered the negative network fields
and cancelled with them. Four minutes later, the positive element
disappeared totally. Dopplergrams in this figure also show that
the granular locations were occupied by the Doppler blue-shift
signals.

\begin{figure}
\centering
\includegraphics[bb=96 260 526
569,clip,angle=0,scale=1.08]{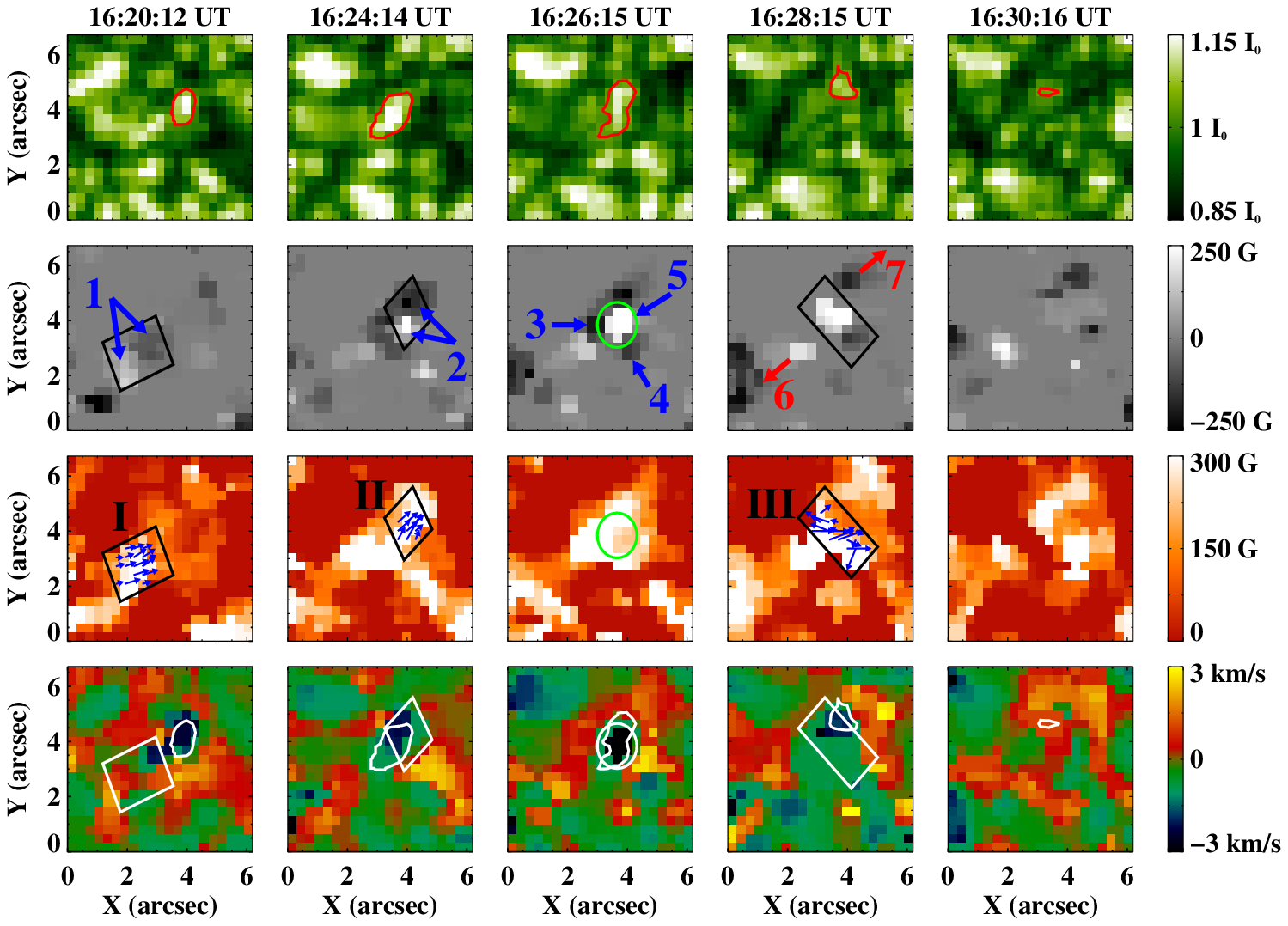} \caption{ Similar to
Fig. 2 but for shrink process of a granule due to the cancellation
between the positive element of a dipole (pointed by a pair of
arrows ``2") and the negative element of another dipole (denoted
by another pair of arrows ``1"). Arrows ``3" and ``4" indicate two
segments dividing from the negative element of dipole ``1", and
arrow ``5" shows the positive element of dipole ``2". Arrows ``6"
and ``7" denote the moving directions of the un-interacting
elements of the two dipoles. The parallelograms and the thin
arrows within them are same as described in Fig. 5. Note that the
arrows in parallelogram ``II" are overplotted from the transverse
fields at the same location at 16:22 UT. The ellipses mark the
position where the remarkable Doppler blue-shifts locate.
\label{fig7}}
\end{figure}

Besides being advected by granular motion, cancelling magnetic
flux also suppresses granular development, as shown in Fig. 7.
With the growing process of a granule, a dipole began to emerge at
16:22 UT, and the horizontal magnetic fields between the dipolar
elements also appeared, with their directions pointing from the
positive element to the negative one, as shown by the arrows in
parallelogram ``II" region. This dipole was much more obvious at
16:24 UT (shown by a pair of arrows ``2"). Before the appearance
of dipole ``2", there already existed another dipole (denoted by a
pair of arrows ``1"). From 16:20 UT to 16:26 UT, the pre-existing
dipole moved to the newly emerged dipole, with an average velocity
of 2.2 km s$^{-1}$. The negative element of dipole ``1" and the
positive element of dipole ``2" cancelled since 16:24 UT. Two
minutes later, the cancelling negative element split into two
segments (denoted by arrows ``3" and ``4", respectively), due to
the collision of the cancelling positive element (see arrow ``5").
Then the positive element of dipole ``1" began to move away from
the cancelling position along arrow ``6" direction with an average
velocity of 2.6 km s$^{-1}$, and finally returned to its birth
place. The negative element of the newly emerging dipole also
moved away from the cancelling position along arrow ``7"
direction, with an average velocity of 2.5 km s$^{-1}$. At 16:28
UT, segment ``3" had disappeared, while segment ``4" and element
``5" were cancelling violently. At this time, strong transverse
fields (inside the parallelogram ``III" region) appeared at the
cancelling position, but their directions were unordered. At 16:30
UT, the two cancelling elements almost disappeared. From the
continuum intensity maps, we can see clearly that the granule
shrank rapidly while the magnetic flux cancellation took place.
Comparing the longitudinal magnetograms with the corresponding
Dopplergrams, we find that, at 16:26 UT, the site of dipolar
element ``5" (outlined by the ellipses) underwent very high
Doppler velocity ($-$3.0 km s$^{-1}$).

\section{Conclusions and discussion}
\label{sect:conclusions and discussion}

By examining 6 typical cases, we present in this paper the
relationship between emerging (cancelling) small-scale magnetic
flux and granular structures in a quiet Sun region near disk
center. A granule develops in a centrosymmetric form in the
condition that no magnetic flux emerges within the granular cell,
and another granular structure develops and splits in a
noncentrosymmetrical form while flux emerges at an outer part of
the granular cell. As soon as a granule breaks up, magnetic flux
emergence as a cluster of mixed polarities appears at the position
of the former granule. A dipole emerges accompanying with the
development of a granule and cancels with pre-existing flux, due
to the advection of the horizontal granular motion. When magnetic
flux cancellation takes place at a granular position, the granule
shrinks and disappears. Our results confirm the idea that granular
flow advects magnetic flux and magnetic flux evolution suppresses
granular development. Furthermore, we uncover the evolution of
transverse fields and changes of Doppler signals during cancelling
process.

Observations show that the horizontal fields of small magnetic
loops are advected toward the surface by the upward motion of the
plasma inside the granules (Lites et al. 1996) and horizontal
motion inside the granules carries the vertical magnetic flux
toward the intergranular lanes (Harvey et al. 2007; Centeno et al.
2007). In this paper, the emergence of the dipole connected by
horizontal fields which pointed from the positive element to the
negative, as shown in Fig. 5, is another case similar to that of
Centeno et al. (2007). However, we cannot exclude the possibility
that the two elements of dipole ``2" spontaneously separated each
other under no driving effect of the granular plasma motion when
they emerged continuously in an $\Omega$-shaped configuration.
Figure 4 shows that magnetic flux emergence as a cluster of mixed
polarities at a position where the granule located was detected as
soon as the granule split. The split of the granule may result
from the emergence of the magnetic flux. There also exists another
possibility that the granule split unaffectedly, and then the flux
emerged upward by magnetic buoyancy or by convection. In
simulation, Cheung et al. (2007) found that the magnetic flux
emergence with a longitudinal flux of more than $10^{19}$ Mx
disturbs the granulation, while small-scale flux tubes with less
than 10$^{18}$ are not sufficiently buoyant to rise coherently
against the granulation and produce no visible disturbance in the
granules. Our results in this observational study are inconsistent
with Cheung et al. (2007), since the emerging magnetic elements
with flux lower than $10^{18}$ Mx, e.g. the event displayed in
Fig. 3, can also affect the granular development.

2 of the 6 cases indicate that magnetic cancellation was triggered
by the pushing effect of the horizontal granular flows (see Figs.
5 and 6). On the other hand, magnetic element also reacted on
granules. The example in Fig. 5 reveals that the development of
the granule was suppressed by the positive element (denoted by
arrow ``3"). At 16:20 UT, the element disappeared; meanwhile the
granule occupied the element position. Vector magnetic field
observations show that the magnetic connection of cancelling
elements was changed, and the transverse fields enhanced during
the cancelling process (see Figs. 5 and 7). These observations
confirm the earlier results that the two components of cancelling
magnetic features are initially not connected by transverse fields
above the photosphere (Zhang et al. 2001). These results are also
consistent with Kubo \& Shimizu (2007), who examined several
collision events and found formation of new magnetic connection.

\begin{figure}
\centering
\includegraphics[bb=24 317 574
617,clip,angle=0,scale=0.8]{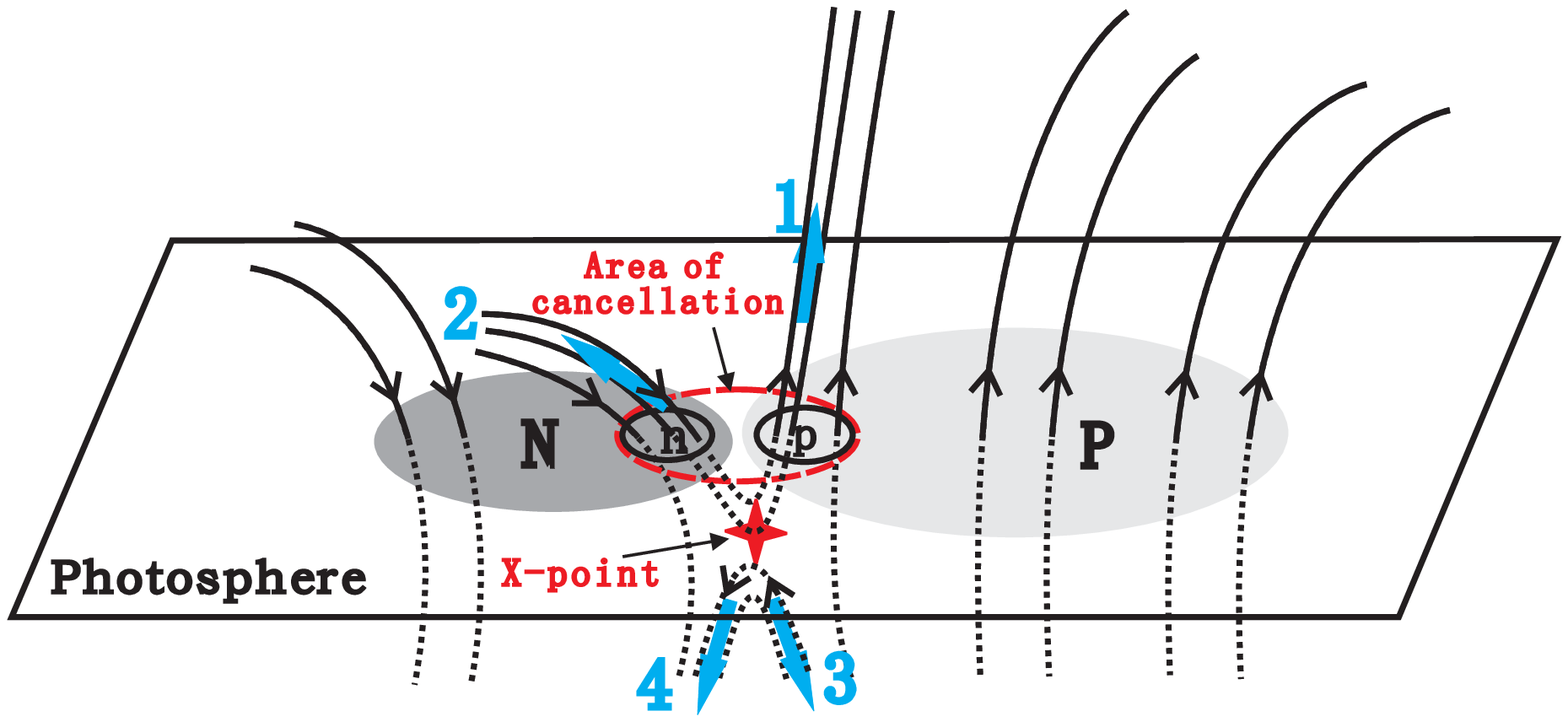} \caption{ Sketch
illustrating formation of large Doppler blue-shifts locating at
one cancelling magnetic element. See the text for details.
\label{fig8}}
\end{figure}

Granules are upwards-moving, hot parcels of gas, exhibiting
blue-shifts in the high-resolution spectral image (Nesis et al.
2001). Not unexpectedly in this study, granules always suffered
Doppler blue-shifts with an average velocity of $-$1.1 km
s$^{-1}$. In Fig. 5, the maximum Doppler blue-shift was $-$2.0 km
s$^{-1}$ at the early emerging stage of dipole ``2", while it
decreased to $-$0.9 km s$^{-1}$ when the dipole well developed. In
Fig. 7, the region with Doppler blue-shifts of about $-$3.0 km
s$^{-1}$ at 16:26 UT is coincident with the site of the cancelling
magnetic element ``5", which belonged to the newly emerging dipole
``2". At the early emerging stage (at 16:20 UT) of this dipole,
when the Doppler blue-shifts are expected to be the largest during
the dipolar lifetime, the Doppler velocity is around $-$2.2 km
s$^{-1}$, smaller than that at cancelling stage. We suggest that
the excess of the blue-shifts at 16:20 in Fig. 7 are produced by
the magnetic flux reconnection below the photosphere, as
demonstrated in Fig. 8. When magnetic reconnection occurs at the
so-called ``X-point", magnetic energy is converted into thermal
energy and kinetic energy. Then bi-directional plasma jets form
and eject from the ``X-point" along the field lines. If the
reconnection takes place below the photosphere, the upward plasma
jets (denoted by arrows ``1" and ``2") move across the solar
surface from inner to outer, and Doppler blue-shifts will be
observed in the photospheric surface. The area where the larger
blue-shifts appear is relevant to the topology of magnetic field
lines. The magnetic field lines jet ``1" moves along are more
vertical than that of jet ``2", so the blue-shifts appear at the
site of one magnetic element (marked by ``p"). Chae et al. (2004)
reported an example of magnetic flux submergence at the flux
cancelling sites. Their observations also revealed that larger
downflows were at the cancelling positive magnetic feature instead
of at the polarity inversion line. If magnetic fields are observed
with lower spatial resolution, small-scale elements cannot be
distinguished separately and several elements combine together to
form a larger one. At this condition, the large blue-shifts will
appear mainly at the adjacent region of the two big cancelling
elements (marked by ``N" and ``P" in Fig. 8). However, we can not
rule out the possibility that the large upward velocities at the
cancellation area are caused by the emerging U$-$shaped flux loops
(Parker 1984; Lites et al. 1995).

Since our this study is limited to only several cases, we will
make a statistical analysis over a large sample of events to
examine whether these results are general or not in our next
study.

\begin{acknowledgements}

We thank the {\it Hinode} team for providing the data. {\it
Hinode} is a Japanese mission developed and launched by ISAS/JAXA,
with NAOJ as domestic partner and NASA and STFC (UK) as
international partners. It is operated by these agencies in
co-operation with ESA and NSC (Norway). This work is supported by
the National Natural Science Foundations of China (G10573025,
40674081 and 40890161), the CAS Project KJCX2-YW-T04, and the
National Basic Research Program of China under grant
G2006CB806303.

\end{acknowledgements}

\label{lastpage}

\end{document}